\documentclass[draft]{agujournal2019}
\usepackage{url} 
\usepackage{lineno}
\usepackage[inline]{trackchanges} 
\usepackage{soul}
\usepackage{graphicx}
\usepackage{amsmath}
\usepackage{amssymb}
\draftfalse

\journalname{Journal of Geophysical Research: Planets}

\begin{document}

%
%


\title{Investigating the extent of bladed terrain on Pluto via photometric surface roughness\thanks{© 2025. All rights reserved.}}

%
%




\authors{I. Mishra\affil{1}, R. Dhingra\affil{1,2}, B. J. Buratti\affil{1}, B. Seignovert\affil{3}, and O. L. White\affil{4,5}}


\affiliation{1}{Jet Propulsion Laboratory, California Institute of Technology}
\affiliation{2}{Applied Semiconductor Material Lithography}
\affiliation{3}{Observatoire des Sciences de l'Univers Nantes Atlantique}
\affiliation{4}{SETI Institute}
\affiliation{5}{NASA Ames Research Center}




\correspondingauthor{Ishan Mishra}{ishan.mishra@jpl.nasa.gov}




\begin{keypoints}
\item We analyze the reflectance behavior of hypothesized bladed terrain regions on the non-encounter side of Pluto using New Horizons data.
\item Photometry can probe effects of surface roughness for scales down to sizes on the order of surface particles (tens of microns in diameter), and help study the poorly-imaged regions of a planet.
\item We find the hypothesized bladed terrain regions to be relatively rough as compared to other areas of Pluto, indicating widespread sublimational erosion of methane-ice on Pluto.
\end{keypoints}

%
%

%
%


\begin{abstract}

NASA's New Horizons spacecraft discovered fields of sub-parallel sets of steep ridges situated in the high-altitude, low-latitude regions in Pluto's encounter hemisphere called ‘bladed terrain’. Thought to be formed due to sublimational erosion of methane ice, bladed terrain represents an active response of Pluto's landscape to current and past climates. The observation of a strong methane signature within the low latitudes of Pluto's non-encounter hemisphere points to the possibility that this terrain type is also present there. To test this hypothesis, in the absence of high resolution images of Pluto's non-encounter hemisphere, we employ photometric analysis of the methane rich regions. We specifically focus on determining the macroscopic surface roughness in selected images, whose photometric-effect can be apparent even in low-resolution images. We employ the `crater-roughness' photometric model of \citeA{buratti_photometry_1985}, which assumes that the surface is covered with parabolic depressions defined by a depth-to-radius ratio parameter $q$ (higher $q$ values correspond to rougher surfaces). Despite the high uncertainty in the retrieved roughness values from our analysis, we can safely conclude that the hypothesized bladed terrain region on the non-encounter hemisphere of Pluto is very rough ($q = 0.47_{-0.11}^{+0.10}$, 2$\sigma$), with the median roughness more than twice that of other broad  regions of Pluto studied in this work, including the encounter-hemisphere bladed terrain region ($q = 0.21_{-0.18}^{+0.08}$, 2$\sigma$).

\end{abstract}

\section*{Plain Language Summary}


NASA's New Horizons spacecraft discovered `bladed terrain', which are bowl-shaped depressions with blade-like spires around the edge that rise several hundreds of meters. Bladed terrain is thought to be formed due to sublimation of methane ice. Although New Horizons could not take high spatial-resolution images of the non-encounter side of Pluto, the spectrometer LEISA found signatures of methane in vast, high-altitude regions here as well, where bladed terrain might also be present. To test this hypothesis, we study the photometry of images of Pluto, i.e, how the intensity of reflected sunlight changes with changing incidence and emission angles of light-rays across various regions on Pluto's surface. We especially focus on the effects of the roughness of a given region, as it characteristically alters its photometric properties. Despite the high uncertainty in the retrieved roughness values from our analysis, we can safely conclude that the hypothesized bladed terrain region on the non-encounter hemisphere of Pluto is very rough, with the median roughness around twice that of the encounter-hemisphere bladed terrain region and six times that of the Sputnik Planitia, the massive basin (or `Pluto's heart') which is visibly devoid of features of the scale of bladed-terrain. 

%
%

%


%
%
%
%

\section*{Introduction}

\begin{figure}[htbp!]
\noindent\includegraphics[width=\textwidth]{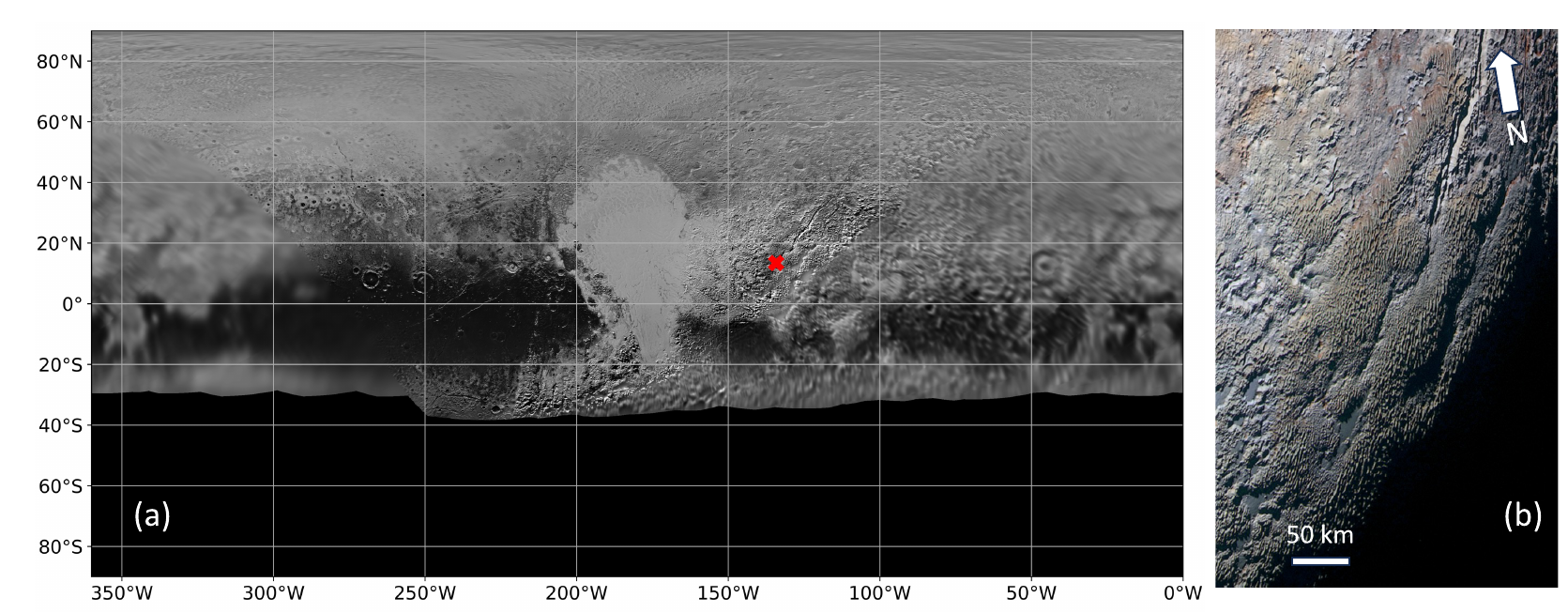}
\caption{(a) Global mosaic of Pluto assembled from from nearly all of the highest-resolution images obtained by the Long-Range Reconnaissance Imager (LORRI) and the Multispectral Visible Imaging Camera (MVIC) on New Horizons \cite{moore_geology_2016}. The mosaic is presented here in equirectangular projection at an equatorial pixel scale of 300 meters per pixel. The red `x' marks the center of the image in panel (b), centered at 134.2°W, 13.4°N (Image credit: NASA/JHUAPL/SwRI). (b) Aligned ridges that form the bladed terrain, as seen in the Tartarus Dorsa region of Pluto's encounter hemisphere. This image (PIA19957) has been adapted from \citeA{moores_penitentes_2017}. It was acquired with MVIC at a spatial resolution of $\sim$ 300 m per pixel, and it is centered at (134.2°W, 13.4°N)}.
\label{fig:bladed_terrain_zoom}
\end{figure}

New Horizons, which encountered Pluto in 2015, discovered fields of roughly evenly spaced, often sub-parallel sets of steep ridges situated on high ground \cite{moore_geology_2016} (Figure \ref{fig:bladed_terrain_zoom}). These landscapes have been labeled as ‘bladed terrain’ and consist of massive deposits of methane \cite{protopapa_plutos_2017, schmitt_physical_2017}, which are observed to occur within 30° latitude of Pluto’s equator and are found almost exclusively at the highest elevations ($>2$ km above Pluto's zero-elevation) \cite{moore_bladed_2018}. These blades are partially analogous to penitentes on terrestrial, low-latitude, high-elevation water-ice fields \cite{moore_geology_2016}, and are thought to be formed by partial erosion of methane deposits over many climate cycles on Pluto \cite{moores_penitentes_2017}. Penitentes are spiked, deep ice deposits that form in high altitude equatorial regions on the Earth, such as in the dry Andes above the elevation of 4 km \cite{lliboitry_origin_1954}, and range in sizes from a few cm to over 5 meters \cite{naruse_preliminary_1997}. Like Earth’s penitentes, the ‘blades’ on Pluto are observed at high altitudes ($> 2$ km), but are much larger in scale, typically spaced 3–7 km crest-to-crest and exhibit relief of $\sim 300$ m from crest to base \cite{moore_bladed_2018}.

Bladed terrain represents an active response of Pluto’s landscape to current and past climates, and is very likely a major terrain type on Pluto \cite{stern_plutos_2021}. \citeA{moores_penitentes_2017} hypothesized the formation of penitentes in the Tartarus Dorsa (Figure \ref{fig:bladed_terrain_zoom}) region of Pluto (110°W to 140°W, 0°N to 20°N) based upon numerical simulations of sublimational erosion of methane ice sheets. They indicate that the penitentes deepen by of order 1 cm per orbital cycle in the present era and grow only during periods of relatively high atmospheric pressure, suggesting a formation timescale of several tens of millions of years, consistent with cratering ages and the current atmospheric loss rate of methane. 

\begin{figure}[htbp!]
\noindent\includegraphics[width=\textwidth]{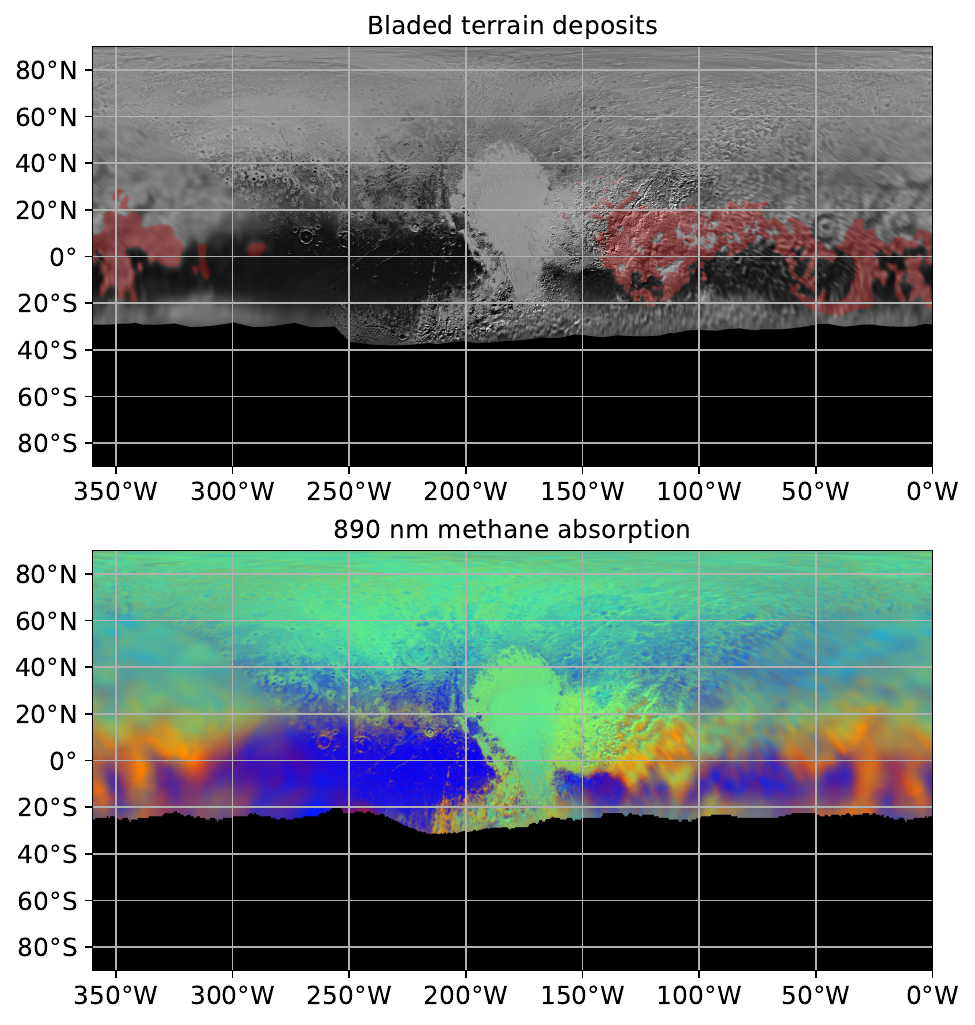}
\caption{\textit{Top:} The Pluto basemap (from Figure \ref{fig:bladed_terrain_zoom}a) is overlain with mapping of the resolved and hypothesized bladed terrain in red (obtained from \citeA{stern_plutos_2021}) , where the resolved bladed terrain regions are in the encounter hemisphere (approx. 100°W - 150°W), and the hypothesized bladed terrain are in the non-encounter hemisphere (approx. 0-100°W and 300-360°W) that correlate with the strong methane absorption and high topography. \textit{Bottom:} Equivalent Width of the 890 nm CH$_4$ absorption band as obtained from New Horizon’s MVIC color observations, overlain on a global mosaic (between 30°S and 90°N) of LORRI and MVIC images. Warmer colors indicate greater CH$_4$ absorption. Map obtained from \citeA{moore_bladed_2018}).}
\label{fig:methane_map}
\end{figure}

There are hints that the observed bladed terrain continues eastward beyond the high resolution coverage of New Horizons’ LOng Range Reconnaissance Imager, or \textit{LORRI} \cite{cheng_long-range_2008}, and the Multispectral Visible Imaging Camera, or \textit{MVIC} \cite{reuter_ralph_2008}, into the poorly resolved non-encounter hemisphere that shows broad areas with high methane absorption in the MVIC data (Figure \ref{fig:methane_map}). \citeA{moore_bladed_2018} present a good correlation between low latitude, high-standing topography where bladed terrain is located, and broad-width methane absorption (centered on the 890 nm absorption band). Moreover, the high methane absorption regions in the non-encounter hemisphere of Pluto also share a similar latitude range as Tartarus Dorsa, where bladed terrain on the encounter-hemisphere are located, being located within 30° of Pluto's equator. The formation of methane deposits in Pluto's equatorial region has also been supported by volatile transport modeling \cite{bertrand_ch4_2019}. Finally, using limb profiles to gauge topography, \citeA{stern_plutos_2021} showed that these non-encounter hemisphere, high methane absorption regions are elevated above surrounding terrain.

LORRI images of the non-encounter side of Pluto do not have the spatial resolution to resolve surface roughness at the scale expected of penitente formation (in the encounter hemisphere image of the bladed terrain, as shown in Figure \ref{fig:bladed_terrain_zoom}, the image resolution is around 300 meters per pixel). In the absence of high resolution images, a promising additional test of whether the bladed terrain extends to the poorly resolved regions of Pluto is to compare the photometric properties, particularly surface macroscopic roughness, of the well resolved bladed terrain regions of Pluto’s encounter hemisphere to regions in the poorly resolved hemisphere that show spectroscopic and topographic hints of bladed terrain. Photometric roughness models peer below the resolution limit of the camera to offer a glimpse of any surface roughness that is in the geometric optics limit \cite{buratti_photometry_1985, helfenstein_geological_1988}.

In this work, we derive photometric roughness of the putative bladed terrain regions on the non-encounter hemisphere of Pluto and compare them to the photometric roughness of the bladed terrain on the encounter hemisphere. Comparison of the retrieved surface roughness parameter values of the known and hypothesized bladed terrain regions may shed light on the extent of this terrain type on Pluto. We also compare the roughness of the bladed terrain regions to other regions (with desirable observation geometry coverage), such as the north polar cap and Sputnik Planitia, to place the values in the context of other geological terrain types and shed light on the diversity of surface roughness on Pluto.  

\section*{Data and Modeling}

\subsection*{Data}

\begin{figure}[htbp!]
\noindent\includegraphics[width=\textwidth]{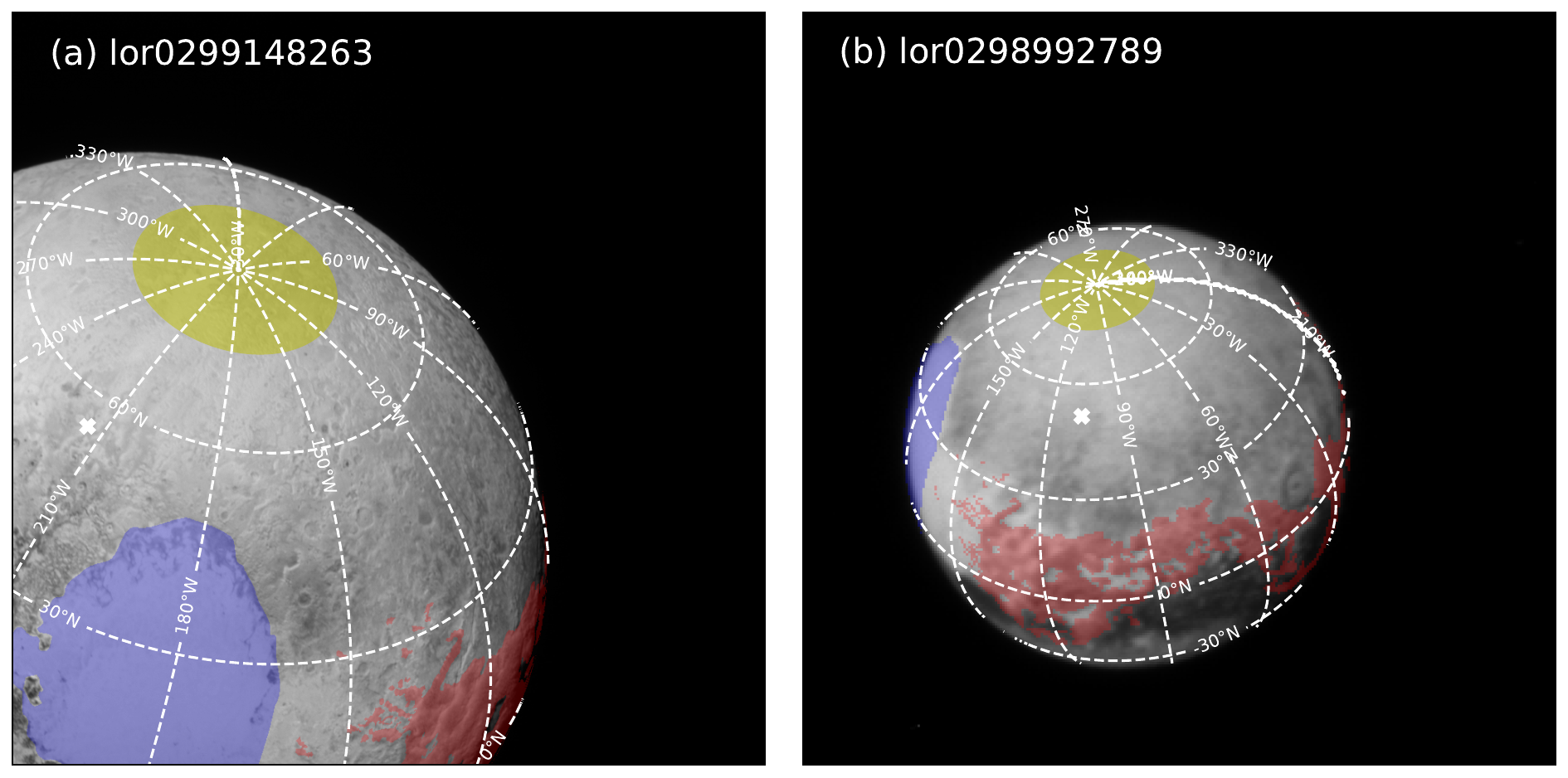}
\caption{The two New Horizons LORRI images, lor0299148263 and  lor0298992789, analyzed in this work, overlain with maps of our regions of interest: the north polar region in yellow (regions north of latitude 75°N), Sputnik Planitia in blue, and the bladed terrain regions in red. A large part of the hypothesized bladed terrain region in the non-encounter hemisphere (approx. between 0-100°W and 300-360°W), as described in Figure \ref{fig:methane_map}, is covered by lor0298992789 (panel (b)) here. We also analyze data from the empty regions of the two images.} The phase angles of observation for lor0299148263 and  lor0298992789 are 16.7° and 15.3° respectively, while the finest spatial resolution is 2.19 km/pixel and 12.82 km/pixel respectively. The white cross in each panel indicates the sub-solar point.
\label{fig:data_images}
\end{figure}

For photometric modeling, we are interested in the variation of reflectance of regions of Pluto’s surface as a function of the observation geometry, particularly emission or incidence angle. Particularly, we looked for images where the collection of pixels provide a wide range in emission and incidence angle coverage in order to reliably study the effect of surface roughness (discussed in the next subsection). We extracted this data from images taken by the New Horizons spacecraft’s LORRI camera, which imaged the near side or the encounter hemisphere of Pluto with very high resolution (resolution $\sim 300$ m/pixel), while the non-encounter hemisphere of Pluto was imaged at a much lower resolution (right-panel of Figure \ref{fig:data_images} has a resolution of 12.82 km/pixel). The non-encounter hemisphere or the non-encounter side of Pluto is approximately the region between 0°- 100°W and 270°- 360°W (Figure \ref{fig:methane_map}) where bladed terrain is also hypothesized to exist \cite{moore_bladed_2018, stern_plutos_2021}. Along with these putative bladed terrain (henceforth PBT) regions, the images we selected cover the bladed terrain on the encounter hemisphere (henceforth EHBT), Sputnik Planitia and the north polar region, to enable comparison of roughness among different geological regions of Pluto (Figure \ref{fig:data_images}). For completeness, we also analyze the leftover/empty regions (regions not in EHBT, Sputnik Planitia or the north polar region) in the two images we ultimately ended up selecting for this analysis, discussed further below.

The data processing pipeline involved two key steps: 1) selecting images that satisfied the observational and geographical constraints described above, for which we used the PDS Ring-Moon Systems Node's OPUS search service tool (\url{https://pds-rings.seti.org/search/}) 2) radiometric calibration of the selected images and incorporating the observation geometry information, for which we used the USGS-ISIS3 (Integrated Software for Imagers and Spectrographs) software \cite{laura_integrated_2023}. The images used in this study are from the New Horizons’ LORRI camera \cite{grundy_new_2007, cheng_long-range_2008}. LORRI is a telescopic panchromatic imager (with no filters) onboard New Horizons, with pivot wavelength of 0.607 $\mu$m. The LORRI images are in ‘.fits’ format on NASA's PDS (Planetary Data System) and before using them we converted the count rate values to radiance values at LORRI’s pivot wavelength (0.607 $\mu$m). The radiance values were then converted to I/F (radiance factor) using the solar flux values at Pluto, since our model calculates reflectance in I/F units as well. Finally, we found that the geometric layers (phase, incidence, emission, latitude, longitude) generated by ISIS3 were offset from the data layer by a significant number of pixels, which had to be manually corrected. 

Only two of the many LORRI image cubes we seived through met our criterion of a wide coverage of emission or incidence angles. Figure \ref{fig:data_images} shows the data from these image cubes with the geological regions of interest in this work. Both the image cubes we selected have a limited range of phase angle ($\sim$ 15° - 17°), so that the effect of diversity in geology/topography of the different regions, which changes the local incidence and emission angles, is highlighted, rather than the effect of the changing phase angle. Further processing of the ISIS generated .cub files, especially to extract the reflectance values from specific regions of an image cube, was performed using a version of the \texttt{galileo-ssi} python package (\url{https://pypi.org/project/galileo-ssi/}) customized for New Horizons/LORRI data. We created masks for specific geological regions of Pluto, along with constraints to limit emission and incidence angles to less than 75° to avoid extreme geometry effects. The I/F values of all pixels from these regions of interest, as a function of emission angle, are our final data product that we fit our roughness model to, as discussed in the next section. We note that in order to make the fitting process computationally tractable, the reflectance curves were binned to intervals of 2° emission angles (see Figure \ref{fig:results}).



\subsection*{The Photometric Roughness Model}

Macroscopic roughness encompasses facets ranging in size from clumps of particles to mountains, craters and ridges. These features alter the specific intensity of a planetary surface in two ways: the local incidence and emission angles are changed by alteration of the surface profile from that of a smooth sphere, and they remove radiation from the scene by casting shadows. Since topographic features at all scales below the resolution of a camera image contribute to photometric roughness, this parameter enables comparison of images of different resolutions, where certain topographic features are resolved in one image but not in the other image.

For inferring roughness from the reflectance data extracted from the regions shown in Figure \ref{fig:data_images}, we use the ‘crater-roughness model’ of \citeA{buratti_photometry_1985} that models the roughness of a surface filled with idealized craters parameterized by a depth-to-radius parameter (first introduced by \citeA{veverka_effects_1972}). The crater-roughness model has been applied to a variety of planetary surfaces, such as comet 19P/Borrelly \cite{buratti_deep_2004}, different terrains on Titan \cite{buratti_titan_2006}, Phoebe \cite{buratti_infrared_2008}, and for high- and low-albedo terrains on Iapetus \cite{lee_roughness_2010}. 

The crater-roughness model calculates the total reflection of a surface covered with parabolic craters as a function of the observation geometry and the ratio of the depth to the radius of the craters. These craters are completely characterized by the shape factor $q$, where $q$ is defined to be:

\begin{equation}
    q = \dfrac{h}{R}
\end{equation}

\noindent where $h$ is the crater depth as measured from the mean surface level, and $R$ is the radius \cite{veverka_effects_1972}. A higher value of $q$ indicates a rough surface, while a lower value of $q$ indicates a smooth surface (with $q = 0$ being a completely smooth surface). While the name of the model includes the word `crater’, it is essentially modeling the effect of any sort of depressions on the surface, whether they are actual craters, or whether they are `valleys’ between the blade ridges. The model calculates reflectance at each grid-point of the crater where the following conditions are met: (1) the point is in the crater, (2) the point is illuminated, (3) the point is visible. The number of points in the crater satisfying these conditions depends crucially on the parameter $q$. 

A general scattering law, which is a combination of the Lunar and Lambert scattering laws, is applied to calculate the reflectance of a crater point:

\begin{equation}
    \dfrac{I}{F}(\mu,\mu_0,\alpha) = A \cdot f(\alpha)\dfrac{\mu_0}{\mu + \mu_0} + (1 - A)\cdot \mu_0
\end{equation}

\noindent where $I$ is the emergent intensity, $F$ is the incident solar flux per unit area per unit solid angle (i.e., $\pi F$ is the total incident flux over a hemisphere), $A$ is the description of the scattering behavior of the surface, $\mu_0$ is the cosine of the incidence angle, $\mu$ is the cosine of emergence angle, and $f(\alpha)$ is the solar phase function of the surface \cite{buratti_voyager_1983}. The scattering parameter $A$ expresses the fraction of reflected radiation that is single scattered: for primarily single scattering behavior, often called Lommel-Seeliger or Lunar-like scattering $A = 1$, while a diffuse or Lambertian scattering behavior implies $A = 0$. In this work, we set A = 0.6 as determined for Pluto by \citeA{buratti_global_2017}. We calculated $f(\alpha)$ for each image using equations 1, 2, and 3, and Figure 2 of \citeA{hillier_characteristics_2021}, deriving $f(\alpha)=1.2$ for the encounter-hemisphere bladed terrain, $f(\alpha)=0.98$ for the non-encounter-hemisphere `putative' bladed terrain, $f(\alpha)=1.38$ for Sputnik Planitia, and $f(\alpha)=1.24$ for the north polar region.

\begin{figure}[htbp!]
\noindent\includegraphics[width=\textwidth]{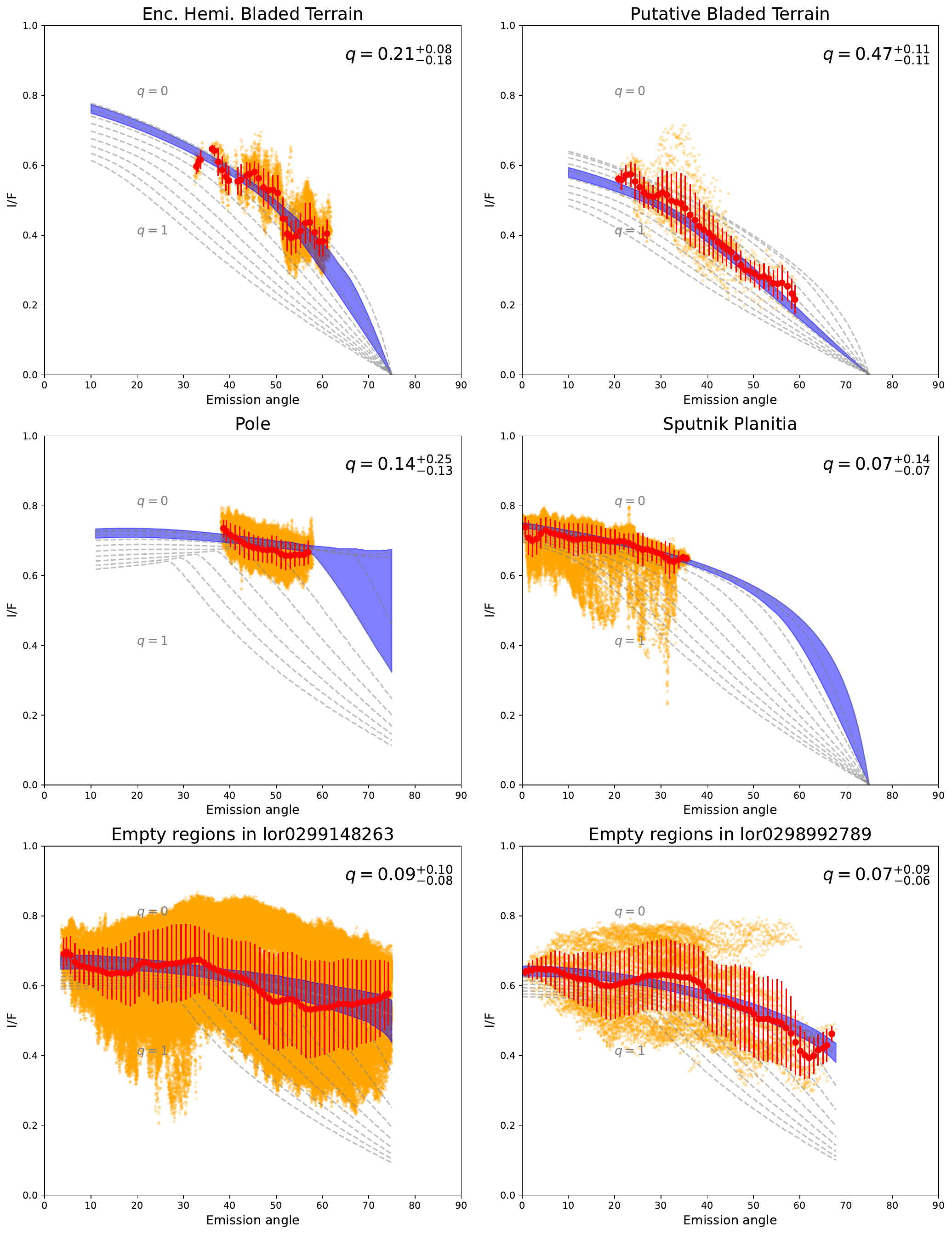}
\caption{Photometric roughness model fits to the reflectance v/s emission angle data of selected regions of Pluto. The orange dots show the reflectance data extracted from the aforementioned regions (see Figure \ref{fig:data_images}), while the red dots show data binned to intervals of 2° with the error bars indicating the standard deviation of reflectance values in each bin. The purple-shaded regions are the model solutions with the associated 2$\sigma$ uncertainty. The dashed grey lines show models for a range of $q$ values from 0 (at the top) to 1 (at the bottom), to provide context to the model fits and highlight how $q$ affects the model. The top right text in each panel shows the retrieved $q$ parameter with associated 2$\sigma$ uncertainty, which is also presented visually in Figure \ref{fig:q_results} for side-by-side comparison.}
\label{fig:results}
\end{figure}

The depth-to-radius ratio parameter $q$ is what we are interested in, as higher values of this parameter would imply a rougher, bladed terrain-like surface. The dashed-grey lines in Figure \ref{fig:results} show the effect of the $q$ parameter on reflectance as emission angle (or equivalently incidence angle) changes. A rougher surface (higher $q$ value) would cause a characteristic downwards inflection in its reflectance curve at small emission angle values, as shadows start dominating at less extreme observation geometry. The disk-resolved form of the crater-roughness model  is particularly useful because it relates surface roughness to limb darkening, which occurs as the emission angle changes, rather than due to changes in solar phase angle. 

A major problem with using a disk-integrated solar phase curve to derive roughness, as inferred from models such as \citeA{hapke_bidirectional_1984}, is that roughness is convolved with other effects (such as the single particle phase function) and thus it cannot be derived uniquely \cite{helfenstein_geological_1988, goguen_new_2010}. Subsequent advancements in modeling photometric roughness, such as the work by \citeA{shiltz_alternative_2023}, have addressed and mitigated some of the shortcomings of the Hapke (1984) model by incorporating new methods to disentangle roughness from other photometric parameters. These improvements highlight the importance of refining photometric models to better capture surface complexity and offer valuable insights into planetary roughness. Disk-resolved measurements obtained by spacecraft are far more diagnostic of surface roughness than integral data sets \cite{helfenstein_geological_1988, buratti_pre-impact_2024}. Of course, all models are limited by their idealization of a morphologically complex surface, but as more and more studies are done, comparisons among different objects can be made. Most importantly though, the photometric effects of penitente-like features should be so extreme that a simple model - we are not arguing that ours is simple - will detect them.

For fitting our photometric model to Pluto's reflectance profiles, we employ a Bayesian inference algorithm. A Bayesian inference framework efficiently explores the parameter space of the model and provides posterior distribution(s) of its free parameter(s) \cite{mishra_comprehensive_2021, mishra_bayesian_2021}, accounting for both uncertainties in the data and degeneracy between model parameters, and has been succesfully applied to photometric analysis of airless icy bodies like Europa \cite <e.g.,>[]{belgacem_estimation_2018}. Our fitting analysis uses two free parameters: 1) $q$, the crater depth-to-radius ratio, and 2) a multiplicative factor for the $f(\alpha)$ parameter. The latter was chosen to account for the inherent uncertainties in the determination of the $f(\alpha)$ values for different regions of Pluto derived by \citeA{hillier_characteristics_2021}. We also observed that our model's absolute reflectance levels matched the data better when $f(\alpha)$ was allowed to vary a little. We use the retrieved posterior distributions of $q$ to report a median value and the $\pm 2\sigma$ bounds on it for every reflectance dataset, and to derive a family of solutions within these bounds that help visualize the uncertainty in the model fits (Figure \ref{fig:results}).

\begin{figure}[htbp!]
\noindent\includegraphics[width=\textwidth]{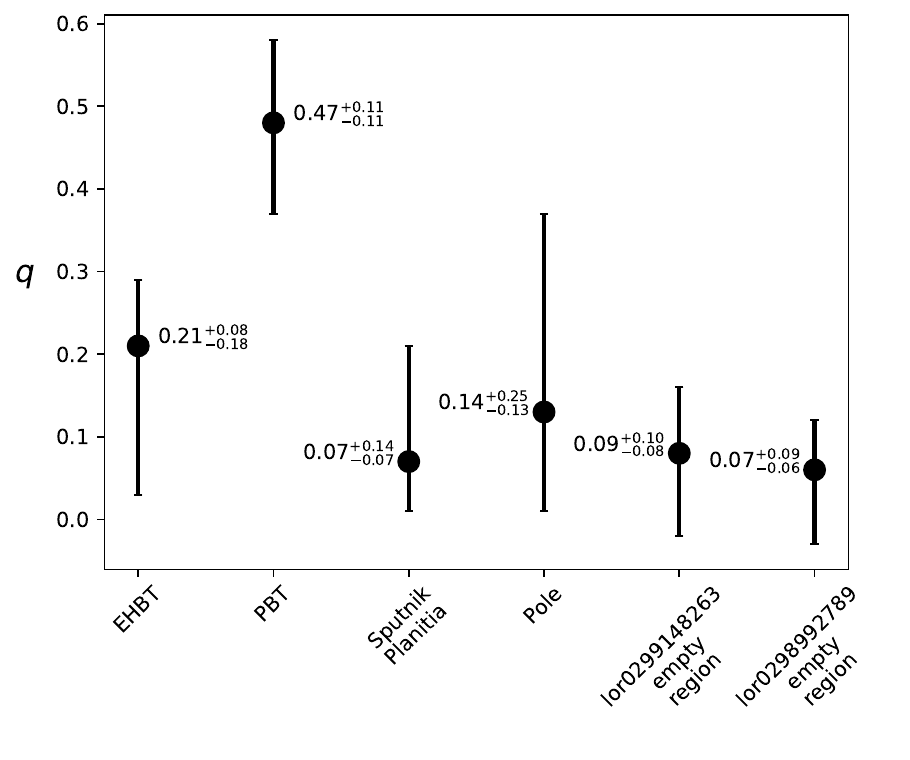}
\caption{Median roughness parameter ($q$) values retrieved for the six regions studied here: encounter hemisphere bladed terrain (``EHBT"), non-encounter hemisphere putative BT (``PBT"), Sputnik Planitia, Pole, and the leftover or empty regions of the two images from which our data comes from (lor0299148263 and  lor0298992789). The error bars correspond to ±2$\sigma$ interval around the median of the retrieved posterior distribution of $q$.}
\label{fig:q_results}
\end{figure}

\section*{Results and Discussion}\label{sec:results}

Figure \ref{fig:results} shows the results of the fit of the crater-roughness model to the data of the four geographical regions shown in Figure \ref{fig:data_images}. The retrieved $q$ parameter values, along with their uncertainties, are visualized in \ref{fig:q_results}.

Firstly, we can ground our derived roughness values in some previous estimates of surface roughness of Pluto. \citeA{moore_geology_2016} provide an estimate of $\sim$ 20° for the V-shaped valleys of the bladed terrain region, via direct inspection of high resolution ($\sim 680$ m/pixel) MVIC image P\_COLOR2 of the Tartarus Dorsa region. Our estimate of the depth-to-radius ratio ($q$) for the encounter-hemisphere bladed region ($q = 0.21_{-0.18}^{+0.08}$) translates to a `mean slope' \cite{hapke_bidirectional_1984} of ($15.64_{-12.75}^{+5.66}$°), which matches well with the ground-truth of $\sim$20° derived by \citeA{moore_geology_2016}. A global estimate of Pluto's roughness  of $\sim 20$° was derived from disk-integrated phase curve data by \citeA{hillier_characteristics_2021}, which matches well with the average value ($\sim 16$° with $\sim 10$° uncertainty) of the roughness of the four regions considered in this work. 

The Sputnik Planitia region is the smoothest (lowest median value of $q$ of 0.07) as compared to the bladed terrain and north polar regions. Simulations show that sublimational erosion of high-altitude ice can create rough topography \cite{moores_penitentes_2017}. However, Sputnik Planitia is 3 to 4 km (1.9-2.5 miles) lower in elevation than most of Pluto's surface \cite{stern_pluto_2018}, so we would not expect sublimational erosion driven roughness to be a factor there, keeping its surface smooth of features at the scale of bladed-terrain (Figure \ref{fig:methane_map}). The north polar region of Pluto is also visibily smooth, and we derive a low roughness value of ({$q = 0.14_{-0.13}^{+0.25}$}) for it. However, the high uncertainties in the retrieved parameter means that the values for Sputnik Planitia and the north polar region are are within 2$\sigma$ 
of the inherently rougher bladed terrain on the encounter hemisphere ($q = 0.21_{-0.18}^{+0.08}$), as shown in Figure \ref{fig:q_results}. The analysis of the empty regions reveals similarly low roughness values of $q = 0.09_{-0.08}^{+0.10}$ and $q = 0.07_{-0.06}^{+0.09}$, further supporting the distinction between the inherently rough bladed terrains and other, smoother regions on Pluto’s surface.

While \citeA{protopapa_disk-resolved_2020} noted that the limited phase angle range of the New Horizons data makes constraining the mean roughness challenging in the Hapke photometric model due to parameter degeneracies \cite{helfenstein_geological_1988}, our approach uses the \citeA{buratti_photometry_1985} crater-reflectance model, which is less complex and primarily sensitive to the roughness parameter. This simpler approach reduces degeneracies and allows us to robustly estimate surface roughness from the available data. Thus, our results are consistent with prior estimates and demonstrate the utility of alternative photometric models in addressing limitations noted by \citeA{protopapa_disk-resolved_2020}.

While the uncertainty in the retrieved roughness value for Sputnik Planitia is high, there are hints of two sub-populations of data points that follow different roughness profiles (orange dots in the bottom-left panel in Figure \ref{fig:results}): one with high reflectance values and another with low reflectance values, roughly divided by the bottom-most dotted-line plot which represents the $q=1$ model (high roughness). The higher reflectance population data, which come from the smooth, central part of Sputnik Planitia (see Figure \ref{fig:methane_map}), follow a shallower curve which correspond to models with smaller (or smoother) $q$ values. In contrast, the lower reflectance population data, which come from the deeply and lighted pitted terrain on the edges of Sputnik Planitita, follow a steeper trend which corresponds to larger (or rougher) $q$ values. Hence, the roughness derived here for the whole population of Sputnik Planitia data averages the roughness of both its smooth and rough regions. This observation is consistent with the findings of \citeA{stern_new_2021}, who identified extensive pitted terrains along the edges of Sputnik Planitia that are associated with higher surface roughness. These pits, formed in volatile ices, contribute to the rougher texture observed in these regions. The central part of Sputnik Planitia is also particularly rich in both N$_2$ and CO \cite{grundy_surface_2016}, two of the most dominant volatiles across Pluto's surface. The high reflectance or albedo of the central region of Sputnik Planitia, along with the dominance of N$_2$ and CO perhaps indicates an area of active convective recycling \cite{moore_geology_2016} that could also result in smooth topography, as we observe in our results. We also acknowledge that the limited phase angle range of the New Horizons data, with no emission angles $>35$°, constrains our ability to fully capture the roughness variations across Sputnik Planitia, as also noted by \citeA{stern_new_2021}.

The uncertainty on the retrieved $q$ parameter value is high for all regions except PBT (see Figure \ref{fig:q_results}). The PBT region spans a wide area, approximately between 0-100°W and 300-360°W in the non-encounter hemisphere (Figure \ref{fig:methane_map}) giving us the most amount of data points spanning a wide range of emission angles, which results in a tighter constraint on the $q$ parameter. The PBT regions are also the roughest of all the regions we have analyzed ($q = 0.48_{-0.11}^{+0.10}$). This could be because the PBT might be inherently rougher than the EHBT regions. Another possibility is that our model underestimates the roughness of the EHBT regions, because of their higher albedo \cite{buratti_global_2017} that could be diluting shadows due to multiple scattering of photons. None of the photometric roughness models currently in literature account for dilution of shadows due to multiply scattered photons, which causes the model to under-estimate roughness, as demonstrated in lab experiments \cite <e.g.,>[]{buratti_photometry_1985}. 

To assess whether the reflectance data for the PBT region are distinct from other regions on Pluto, we performed pairwise Kolmogorov-Smirnov (KS) tests between the reflectance distributions of the PBT region and those of other regions, including Sputnik Planitia, the North Pole, EHBT, and the two empty regions. The KS test is a nonparametric method that compares the cumulative distribution functions (CDFs) of two datasets, quantifying the maximum difference between them. The resulting KS statistic and p-value provide a measure of whether the two datasets are drawn from the same distribution, with a p-value below 0.05 indicating significant differences.

\begin{figure}[htbp!]
\noindent\includegraphics[width=\textwidth]{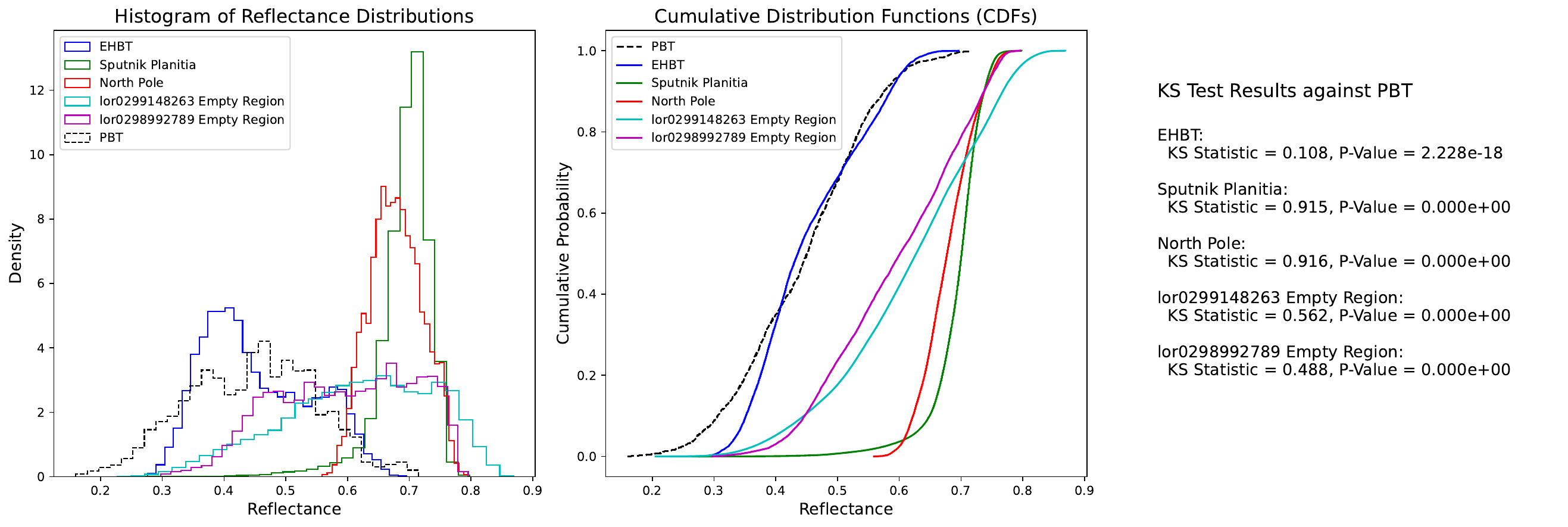}
\caption{Comparison of reflectance data distributions and cumulative distribution functions (CDFs) for the putative bladed terrain (PBT) region and other regions on Pluto. (Left) Histograms of reflectance values for the PBT region and comparison regions, shown as outlined curves for clarity. The PBT region is represented with a dashed line. (Center) CDFs of reflectance values for the PBT region and comparison regions, highlighting differences in their cumulative probabilities. (Right) Summary of Kolmogorov-Smirnov (KS) test results, including KS statistics and p-values for pairwise comparisons between the PBT region and other regions. The KS test indicates that the PBT region is statistically distinct from all other regions (p-value $<0.05$), although the histograms and CDFs indicate that it shares similar reflectance characteristics with the encounter hemisphere bladed terrain (EHBT).}
\label{fig:ks_test}
\end{figure}

The results of the KS tests (summarized in Figure \ref{fig:ks_test}) reveal two key insights. First, the PBT reflectance data are statistically distinct (p $<0.05$) from all other regions tested, indicating that the unique high surface roughness derived for the PBT region is indeed tied to a unique reflectance distribution. Second, while the PBT reflectance data differ significantly from most other regions, their histograms and CDFs are remarkably similar to those of the EHBT region. This similarity suggests that the PBT region shares similar albedo characteristics with EHBT, despite its slightly higher derived roughness.

These results support the interpretation that the PBT region's roughness is an intrinsic property of the terrain rather than an artifact of external factors, such as photometric conditions. The similarity in albedo between PBT and EHBT further reinforces the hypothesis that the PBT region represents a bladed terrain type, consistent with previously identified EHBT regions. Future studies could investigate whether other photometric or compositional properties corroborate this connection.

The key result from our work is that the roughness of the methane-rich regions in the non-encounter hemisphere of Pluto is quite high, and provides support to the hypothesis that these high-altitude, methane-rich regions could also be covered with bladed-terrain like the encounter hemisphere (Figure \ref{fig:bladed_terrain_zoom} \cite{moore_bladed_2018}. Along with methane signatures on the non-encounter hemisphere, isolated outcrops of bladed-terrain-like sharp units have also been identified with limb topography \cite{stern_plutos_2021}. Putting these together with the high roughness value derived in our work, we support the idea that Tartarus Dorsa (Figure \ref{fig:bladed_terrain_zoom}) forms the western extreme of a vast belt of bladed terrain deposits extending across the planet, within an equatorial zone ($\lesssim 30$° latitude) spanning $> 220$° of longitude, primarily on the non-encounter hemisphere of Pluto. However, we cannot rule out the possibility that the high-roughness of the methane-rich areas in the non-encounter hemisphere could be caused by landforms other than bladed terrain, although this hasn't been suggested in literature. Another caveat in our results is that we are assuming that roughness at scales much smaller than the bladed-terrain (such as clumps of particles) do not cause the variation we see between the EHBT regions and PBT in the non-encounter hemisphere. However, the processes that could cause variation at that scale, such as impact gardening, micro-meteoritic bombardment, haze infall, etc. \cite{hillier_characteristics_2021} should not vary globally to a significant extent, making larger features on the scale of bladed terrain the dominant source of roughness. A broader analysis of roughness across all of Pluto's surface can reveal if high-roughness values are restricted to methane rich regions or not, but that is beyond the scope of this work.

The presence of these bladed terrains on Pluto provides important insights into the geology and dynamics of Pluto's surface. Rough terrain extend all across the non-encounter hemisphere of Pluto and have likely formed via sublimation erosion of methane ice. The youthful bladed-terrain could be the most dominant geological landform on Pluto. The analysis presented in this work can be extended to other icy bodies in the outer solar system, such as Europa, where penitentes-like structures have been hypothesized to exist as well \cite{hobley_formation_2018}. Determining the diversity of roughness of planetary surfaces adds to our understanding of the processes that shape icy bodies in our solar system, and highlights the complexity and diversity of planetary surfaces.

\section*{Conclusions}

This study provides a comprehensive analysis of surface roughness across several geographical regions of Pluto using photometric modeling. The findings reveal that the putative bladed terrains (PBT) in the non-encounter hemisphere, which New Horizons could only image at low resolution, exhibit significantly higher roughness compared to other regions across Pluto. The encounter-hemisphere bladed terrains (EHBT) are also rough but less so than the PBT. This difference may be attributed to the higher albedo of EHBT regions, which likely causes multiple scattering of photons and a corresponding underestimation of roughness in photometric models. The pairwise Kolmogorov-Smirnov (KS) tests further demonstrate that the PBT region has statistically distinct reflectance distributions compared to other regions, reinforcing the conclusion that the high roughness of PBT is intrinsic and not merely an artifact of observational or photometric conditions. These results support the hypothesis that the PBT regions are distinct geological features, possibly formed through sublimation-driven erosion of methane ice.

The robust methodology applied here demonstrates consistency with previous estimates of Pluto's surface roughness and underscores the utility of alternative photometric models in planetary surface studies. While the analysis confirms the roughness variations among distinct terrains, future investigations could explore additional compositional or morphological factors influencing surface roughness and extend similar approaches to other icy bodies in the solar system.

\section*{Open Research}

The data analysed in this work were obtained from The PDS Ring-Moon Systems Node's OPUS search service, available at \url{https://pds-rings.seti.org/search/}. Pre-processing of the data cubes downloaded from OPUS was performed using the Integrated Software for Imagers and Spectrometers (ISIS) \cite{laura_integrated_2023} (\url{https://doi.org/10.5281/zenodo.7644616}). The python software used for data analysis and plotting are NumPy \cite{harris2020array}, Jupyter \cite{granger_jupyter_2021}, Matplotlib \cite{Hunter:2007}, SciPy \cite{2020SciPy-NMeth}, and dynesty \cite{speagle_dynesty_2020}.

\acknowledgments

This research was funded by the New Horizons Project, and carried out at the Jet Propulsion Laboratory, California Institute of Technology, under contract to the National Aeronautics and Space Administration. We would like to thank Paul Schenk (LPI) and Jeff Moore (NASA Ames) for producing and sharing the methane absorption map used in Figure \ref{fig:methane_map}.

\bibliography{manucript_accepted}

\end{document}